\documentclass[preprint,prb,aps]{revtex4}
\usepackage{amsmath,amssymb}
\usepackage{graphicx}
\usepackage{dcolumn}
\usepackage{bm}

\begin{document}

\title[Quadratic algebra and nonlinear coherent states]{An observation of
quadratic algebra, dual family of nonlinear coherent states and their non-classical properties,
in the generalized isotonic oscillator}

\author{V~Chithiika Ruby and  M~Senthilvelan}

\affiliation{Centre for Nonlinear Dynamics, School of Physics,
Bharathidasan University, Tiruchirapalli - 620 024, India.}

\begin{abstract}
In this paper, we construct nonlinear coherent states for the generalized isotonic
oscillator and study their non-classical properties in-detail. By transforming the
deformed ladder operators suitably, which generate the quadratic algebra, we
obtain Heisenberg algebra. From the algebra we define two non-unitary and
an unitary displacement type operators. While the action of one of the non-unitary
type operators reproduces the original nonlinear coherent states,
the other one fails to produce a new set of nonlinear coherent states (dual pair).
We show that these dual states are not normalizable.  For the nonlinear
coherent states, we evaluate the parameter $A_3$ and examine the non-classical
nature of the states through quadratic and amplitude-squared squeezing effect.
Further, we derive analytical formula for the $P$-function, $Q$-function and
the Wigner function for the nonlinear coherent states.  All of them confirm the
non-classicality of the nonlinear coherent states. In addition to the above,
we obtain the harmonic oscillator type coherent states from the unitary displacement operator.
\end{abstract}

\pacs{03.65.-w, 03.65.Ge, 03.65.Fd}

\maketitle

\section{Introduction}
In this paper, we study the construction of
nonlinear coherent states and their dual pair for a conditionally solvable
potential \cite{car}
\begin{eqnarray}
V(y) = \left(\frac{m_0 \omega^2 }{2} y^2 + \frac{g_a(y^2 - a^2)}{(y^2 + a^2)^2}\right),
\label{lad1}
\end{eqnarray}
where $m_0, \omega, a$ and $g_a$ are parameters. Introducing ${ y = \sqrt{\frac{\hbar}{{m}_0 \omega}}\; x}$ and choosing the
parameters $a$ and $g_a$ in such a way that ${ a^2 =  \frac{\hbar}{2 m_0 \omega}}$ and
${ g_a = \frac{2 \hbar^2}{m_0}}$ with the energy eigenvalues  $e_n =\hbar \omega E_n$, we
can rewrite the time independent Schr\"{o}dinger equation associated with the above potential is of the form
\begin{eqnarray}
\hspace{-0.8cm}-\frac{1}{2}\frac{d^2 \psi_n(x)}{dx^2}+ \frac{1}{2}\left(x^2 + \frac{8(2x^2 - 1)}{(2x^2 +1)^2}\right)
\psi_n(x) =  E_n \psi_n(x).
\label{lad3b}
\end{eqnarray}

On solving Eq. (\ref{lad3b}), the authors of Ref. \onlinecite{car} obtained the eigenfunctions and
energy eigenvalues as
\begin{eqnarray}
\psi_n(x) &=& {\cal N}_n \displaystyle{\frac{{\cal P}_n(x)}{(1+2x^2)}} e^{\displaystyle-{x^2}/2},
\label{lad3c}\\
E_n &=& -\frac{3}{2} + n, \quad \qquad \qquad n = 0,3,4,5,...,
\label{lad3a}
\end{eqnarray}
where the polynomial factors ${\cal P}_n(x)$ are given by
\begin{eqnarray}
\qquad \qquad {\cal P}_n(x) = \left\{\begin{array}{c}
\hspace{-1.2cm} 1, \quad \qquad \qquad \qquad \qquad \qquad \qquad \qquad\quad \mbox{if}\;\;n = 0 \\
  H_n(x) + 4 n H_{n-2}(x) + 4 n (n-3) H_{n-4}(x), \;\mbox{if}\;\;n = 3, 4, 5,...\\
  \end{array}\right.
\end{eqnarray}
and the normalization constant is given by
\begin{eqnarray}
{\cal N}_n =\left[\displaystyle{\frac{(n-1)(n-2)}{2^n n! \sqrt{\pi}}}\right]^{1/2}, \quad n = 0,3,4,5,....
\label{lad4}
\end{eqnarray}
We consider Eq. (\ref{lad3b}) as the number operator equation after subtracting
the ground state energy ${\displaystyle E_0 = -\frac{3}{2}}$ from it, that is
\begin{eqnarray}
\hspace{-1cm}\hat{N}_0|n\rangle = n|n\rangle.
\label{lad5}
\end{eqnarray}

From the ground state solution $\psi_0$, we can identify a superpotential for (\ref{lad1}) of the form
\begin{equation}
\hspace{-1.5cm}\phi = -\frac{\psi^{'}_0}{\psi_0} = x + \frac{4 x}{1 + 2 x^2}.
\label{susy}
\end{equation}

During the past few years, attempts have been made to analyze the system (\ref{lad3b})
in different perspectives.  In the following we briefly summarize these activities.
To start with Fellows and Smith have shown that the solvable potential given in Eq. (\ref{lad3b})
is a supersymmetric partner potential of the harmonic oscillator \cite{fellows}.
Kraenkel and Senthilvelan have considered
the exactly solvable potential of Eq. (\ref{lad3b}) and obtained a different class of exactly
solvable potentials by transforming the Schr\"{o}dinger equation (\ref{lad3b}) into the position-dependent mass
Schr\"{o}dinger equation and solving the underlying equation \cite{sen}.
The coherent states for the position-dependent mass Schr\"{o}dinger equation were then
constructed  by the present authors with an illustration of the above exactly solvable
nonlinear oscillator potential \cite{chi1}.
Sesma considered the Schr\"{o}dinger equation associated with potential given in Eq. (\ref{lad1}) and transformed
the equation to the confluent Heun equation and solved the later equation numerically \cite{sesma}.
For certain parametric values the author has given
quasi-polynomial solutions.  The present authors have constructed various non-classical states including intelligent states,
nonlinear coherent states and even and odd nonlinear coherent states for this nonlinear oscillator
 \cite{chi2}.  In a very recent paper, Saad {\it{et al}} \cite{hall2}
have generalized the potential given in  Eq. (\ref{lad1}) and solved
the associated Schr\"{o}dinger equation with the help of asymptotic iteration method and derived the
quasi-exact solutions of Sesma as well as the results of Cari\~{n}ena {\it{et al}}  as
special cases. The $N$-dimensional version of the potential given in Eq. (\ref{lad1}) has also been solved using supersymmetry
technique in Ref. \onlinecite{hall1}. In a very recent work, we have also constructed squeezed states and 
nonlinear squeezed states for the nonlinear system (\ref{lad3b}) \cite{chi3}. 

We note here that through supersymmetry technique Junker and Roy \cite{proy} have derived a class of
conditionally exactly solvable potentials, including the potential given in Eq. (\ref{lad3b}), which are supersymmetric partners of the linear harmonic oscillator.  The symmetry algebra for the conditionally
exactly solvable potentials had also been discussed briefly in Ref. \onlinecite{proy}. Very recently
Quesne has come up with a more general solvable form of inverse square type potential of the form (\ref{lad1}) \cite{quesne_iso}.
Using point canonical transformation method the author had shown that the bound state solution
to the Schr\"{o}dinger equation can be expressed in terms of Laguerre exceptional orthogonal polynomials.
Interestingly this general inverse square type potential admits shape invariant property unlike the potential (\ref{lad1}).
The parameter which makes the general inverse square potential as shape invariant one is
being fixed in the potential (\ref{lad1}) and consequently the shape invariance property is lost by (\ref{lad1}).

In our earlier study, from the solution of the Schr\"{o}dinger equation, we have constructed
deformed annihilation $\hat{N}_{-}$ and creation $\hat{N}_{+}$ operators  of the
form
\begin{eqnarray}
\hspace{-1cm}\sqrt{2}\hat{N}_{-} &=& \left[\frac{2(2x^2 - 1)}{(1+2x^2)^2}-1\right]\left[\frac{d}{dx} + \phi\right]+\left[\frac{d}{dx} + x \right]\hat{N}_{0} ,
\label{lad8}
\\
\hspace{-1cm}\sqrt{2}\hat{N}_{+} &=& -\frac{2(2x^2 - 1)}{(1+2x^2)^2}\left[\frac{d}{dx} + \phi\right]+\left[-\frac{d}{dx} + x \right]\hat{N}_{0}.
\label{lad9}
\end{eqnarray}

We note here that one can express these deformed annihilation and creation operators
in the form $\hat{A}^{\dag}\hat{a} \hat{A}$ and $\hat{A}^{\dag}\hat{a}^{\dag} \hat{A}$ by taking into
account the fact that the potential given in  Eq. (\ref{lad3b}) is a partner potential of the harmonic oscillator
\cite{berger}.  Here, $\hat{a}$ and $\hat{a}^{\dag}$ are annihilation and creation operators of the
harmonic oscillator potential and $\hat{A}$ and $\hat{A}^{\dag}$ are the operators which relate
the potential given in  Eq. (\ref{lad3b}) to the harmonic oscillator
and are given by
\begin{eqnarray}
\hat{A} = \frac{1}{\sqrt{2}}\left(\frac{d}{dx} + \phi\right),\qquad \quad \;\;\;
\hat{A}^{\dagger} =  \frac{1}{\sqrt{2}}\left(-\frac{d}{dx} + \phi \right)
\label{op}
\end{eqnarray}
with $\phi$ is given in Eq. (\ref{susy}).

One can use these deformed ladder operators $\hat{N}_{-}$ and  $\hat{N}_{+}$ to generate
deformed algebras.  Using the framework given in Ref. \onlinecite{Bambah}, in this paper, we construct coherent states from the
deformed quadratic algebra.  From the deformed $su(1,1)$ algebra we define two non-unitary displacement type
operators, namely $\hat{D}(\alpha)$ and $\hat{\widetilde{D}}(\alpha)$. 
In this work, we demonstrate that while the action of one of them on the ground
state (in our case $|3\rangle$) reproduces the original nonlinear coherent states, which are reported in
our earlier paper \cite{chi2}, the other one fails to produce
a new set of nonlinear coherent states which are known as their dual pair \cite{dual}.
The aim of this paper is make a comprehensive study on the non-classical
properties exhibited by the nonlinear coherent states.  In particular, we investigate the
parameter $A_3$ and examine non-classical nature
of the nonlinear coherent  states by evaluating quadrature squeezing and amplitude-squared squeezing.
Further, we derive the analytical expressions for the $P$-function, the $Q$-function and the Wigner function
for the non-classical states.  The higher order singularity of the $P$-function and the
partial negativity of the Wigner function do confirm the non-classical properties
of the nonlinear coherent states. All these informations about this
system are new to the literature.  From the unitary displacement operator
we obtain usual form of canonical coherent states.  In the following, we mention some possible 
applications and extensions that can be carried out from 
the nonlinear coherent states constructed in this work. 

Our investigations confirm that the coherent states of (\ref{lad3b}) do 
admit Poissonian distribution only. These non-Gaussian states, which cannot be 
reduced to Gaussian wave packets in position and momentum coordinates, are shown to
possess classical/non-classical properties. These non-classical non-Gaussian states
can be used to construct Schr\"{o}dinger's cat states through suitable superpositions 
\cite{cat, tav}. These Schr\"{o}dinger's cat states are shown to be useful  resources for fault-tolerant 
quantum information processing, see for example Refs. \onlinecite{qip} and \onlinecite{nature}. Recently attention 
 has also been paid to construct other types of non-classical
states such as photon-modulated coherent and nonlinear coherent states
since these coherent states are come through by excluding some Fock states 
\cite{pm}.  Finally, we note that the study of nonlinear time evolution  
of the non-classical non-Gaussian states are of contemporary interest 
due to the cubic nonlinearity admitted by the function $f(n)$ \cite{rev, Manko} since 
at least third order nonlinear operations are necessary for building a universal quantum computer \cite{nature}.

We organize our presentation as follows. In the following section, we discuss the method of obtaining Heisenberg algebra
from the deformed annihilation and creation operators.
In section 3, we present the construction of nonlinear coherent states and their dual pair  from the Heisenberg algebra
for this nonlinear oscillator. Consequently, we analyze certain photon statistical properties, quadrature and amplitude-squared squeezing properties exhibited by the nonlinear coherent states and the harmonic oscillator type coherent states in section 4.
Followed by this, in section 5,
we study the quadrature distribution and quasi-probability distributions for the nonlinear coherent states.
Finally, we present our conclusions in section 6.

\section{\bf Deformed oscillator algebra}
\label{sec2}
The deformed ladder operators $\hat{N}_{-}$ and $\hat{N}_{+}$ satisfy the relations \cite{chi2}
\begin{eqnarray}
\hat{N}_{-}|n\rangle &=& \sqrt{n}\;f(n)\;|n-1\rangle,
\label{lad9a} \\
\hat{N}_{+}|n\rangle &=& \sqrt{n+1}\;f(n+1)\; |n+1\rangle, \quad n = 0, 3, 4, 5,...
\label{lad10}
\end{eqnarray}
with $f(n) = \sqrt{(n-1)(n-3)}$. Since $f(n)$ has zeros at $n = 1$ and $3$, we relate
the annihilation ($\hat{a}$) and creation operators ($\hat{a}^{\dagger}$)
to the deformed ladder operators $\hat{N}_{-}$ and $\hat{N}_{+}$ through the relations,
\begin{eqnarray}
\hat{a} = \frac{1}{\hat{f}(\hat{N}_0 + 1)} \hat{N}_{-}, \qquad
\hat{a}^{\dagger} =\frac{1}{\hat{f}(\hat{N}_0)} \hat{N}_{+} , \quad n = 0, 3, 4, 5,...
\label{lad10a}
\end{eqnarray}
in which we preserve the ordering of operators $\hat{f}(\hat{N}_{0}),\; \hat{N}_{-}$ and $\hat{N}_{+}$.
Specifically the operators $\hat{a}$ and $\hat{a}^{\dagger}$ act
on the states $|0\rangle$ and $|3\rangle$ yield
\begin{eqnarray}
\hat{a}|0\rangle &=& \frac{1}{\hat{f}(\hat{N}_{0}+1)} \hat{N}_{-}|0\rangle = 0, \qquad
\hat{a}^{\dagger}|0\rangle = \frac{1}{\hat{f}(\hat{N}_{0})} \hat{N}_{+}|0\rangle = 0, \\
\hat{a}|3\rangle &=&  \frac{1}{\hat{f}(\hat{N}_{0}+1)} \hat{N}_{-}|3\rangle =  0, \qquad
\hat{a}^{\dagger}|3\rangle =  \frac{1}{\hat{f}(\hat{N}_{0})} \hat{N}_{+}|3\rangle = \sqrt{4}\;|4\rangle.
\label{lad10c}
\end{eqnarray}
For the remaining states, the operators produce
\begin{eqnarray}
\hat{a}|n\rangle &=& \sqrt{n}\;|n-1\rangle, \\
\hat{a}^{\dagger}|n\rangle &=& \sqrt{n+1}\;|n+1\rangle, \quad n = 4, 5, 6, 7, ...
\label{lad10b}
\end{eqnarray}
and $\hat{N}_{0} = \hat{a}^{\dagger} \hat{a}$.

From (\ref{lad9a}) and (\ref{lad10}), we observe that $\hat{N}_{-} |0\rangle = 0$
and $\hat{N}_{+}|0\rangle = 0$.  As a consequence
the ground state can be considered as an isolated one.  Further, the expression $\hat{N}_{-}|3\rangle = 0$ implies
that the first excited state $|3\rangle$ acts as a
ground state. This is due to the fact that $f(n)$ has
zeros at $n = 1$ and $3$. Because of this fact, the Hilbert space ${\cal H}$
consists of states $|0\rangle, |3\rangle, |4\rangle,...$ splits up into two invariant
sub-spaces, namely (i) $|\Psi\rangle = |0\rangle$ and (ii) $|\Psi^{'}\rangle = \sum^{\infty}_{n = 3} c_{n} |n\rangle$ for the
operators $\hat{N}_{-}$ and $\hat{N}_{+}$  \cite{Manko}. We consider the sub-Hilbert space, ${\cal H'}$, spanned
by the eigenstates, $|3\rangle, |4\rangle, |5\rangle,...$ and exclude the ground state $|0\rangle$ for further discussion.

The operators $\{\hat{N}_{-}, \hat{N}_{+},\hat{N}_{0}\}$ satisfy the following quadratic algebra
\cite{ proy, Bambah, quadratic}
\begin{eqnarray}
[\hat{N}_{+},\hat{N}_{-}]|n\rangle = [5\hat{N}_{0} - 3 \hat{N}^2_{0}]|n\rangle, \qquad
[\hat{N}_{0}, \hat{N}_{\pm}]|n\rangle = \pm \hat{N}_{\pm} |n\rangle.
\label{lad11}
\end{eqnarray}

The deformed $su(1,1)$ algebra given in Eq. (\ref{lad11}) is equivalent to the one studied by Delbecq and Quesne in Ref. \onlinecite{quesne1993}.
In the present case we observe that one of the deformed functions is unity and the other one is a quadratic polynomial in $\hat{N}_0$.
The above algebra has the Casimir operator of the type \cite{quesne1993}
\begin{eqnarray}
\hat{C} = \hat{N}_{-}\hat{N}_{+} + \hat{h}(\hat{N}_0) = \hat{N}_{+}\hat{N}_{-} + \hat{h}(\hat{N}_0 - 1),
\label{lad12}
\end{eqnarray}
where $\hat{h}(\hat{N}_0)$ is a real function which involves only $\hat{N}_0$.
Following the ideas given in Ref.  \onlinecite{quesne1993} we
find  $\hat{h}(\hat{N}_0)$ is of the form
\begin{eqnarray}
\hat{h}(\hat{N}_0) = \frac{5}{2}\hat{N}_0(\hat{N}_0+ 1)- \hat{N}_0 (\hat{N}_0+1)(\hat{N}_0 + \frac{1}{2}).
\end{eqnarray}

To construct nonlinear coherent states \cite{Manko, tavanon, Gbook} for the system (\ref{lad3b}),
we transform the operators $\hat{N}_{-}$ and $\hat{N}_{+}$ in such a way that they
satisfy the Heisenberg algebra. By transforming either one of the operators  suitably or
transforming both operators simultaneously we can achieve  this goal.
In the following, we consider all these three possibilities. To begin with we
rescale these deformed operators suitably and obtain the desired Heisenberg algebra.
We then construct the nonlinear coherent states using the transformed ladder
operators.

First let us rescale $\hat{N}_{+}$ such that \cite{Bambah}
\begin{eqnarray}
\hspace{-1.5cm}\hat{{\cal N}}_{+} = \hat{N}_{+} \hat{F}(\hat{C},\hat{N}_0),
\label{lad13}
\end{eqnarray}
where $\hat{{\cal N}}_{+}$ is the new deformed ladder operator and
$\displaystyle{\hat{F}(\hat{C}, \hat{N}_0) = \frac{\hat{N}_0 + \delta}{\hat{C}-\hat{h}(\hat{N}_0)}=\frac{\hat{N}_0 + \delta}{\hat{N}_{-}\hat{N}_{+}}}$, where $\delta$ is a parameter.

We can generate Heisenberg algebra for the nonlinear system (\ref{lad3b}),
through the newly deformed ladder operator given in Eq. (\ref{lad13}),  in the form
\begin{eqnarray}
 \mbox{Case: (i)}\;\;\; [\hat{N}_{-}, \hat{{\cal N}}_{+}]|n\rangle = |n\rangle, \;\; [\hat{N}_{1}, \hat{N}_{-}]|n\rangle = - \hat{N}_{-}|n\rangle, \;\;\;
[\hat{N}_{1}, \hat{\cal N}_{+}]|n\rangle = \hat{\cal N}_{+}|n\rangle,
\label{lad14}
\end{eqnarray}
where $ \hat{N}_{1} = \hat{\cal N}_{+} \hat{N}_{-}$ is a number operator in the sub-Hilbert space
spanned by the states $|3\rangle, |4\rangle, |5\rangle,...$.

Similarly by rescaling the ladder operator $\hat{N}_{-}$ such that
\begin{eqnarray}
\hat{\cal N}_{-} = \hat{F}(\hat{C}, \hat{N}_0) \hat{N}_{-},
\label{lad13b}
\end{eqnarray}
where $\hat{\cal N}_{-}$ is the new deformed ladder operator with
$\displaystyle{\hat{F}(\hat{C}, \hat{N}_0) = \frac{\hat{N}_0 + \delta}{\hat{C}-\hat{h}(\hat{N}_0)}=\frac{\hat{N}_0 + \delta}{\hat{N}_{-}\hat{N}_{+}}}$, we can generate the second set of Heisenberg algebra in the form
\begin{eqnarray}
 \mbox{Case: (ii)}\;\;\;  [\hat{\cal N}_{-}, \hat{N}_{+}]|n\rangle = |n\rangle, \;\; [\hat{N}_2, \hat{\cal N}_{-}]|n\rangle = -\hat{\cal N}_{-}|n\rangle, \;\;\;
[\hat{N}_2, \hat{N}_{+}]|n\rangle = \hat{N}_{+}|n\rangle,
\label{lad15}
\end{eqnarray}
where $\hat{N}_2 = \hat{N}_{+}\hat{\cal N}_{-}$ is also a number operator in the sub-Hilbert space
spanned by the states $|3\rangle, |4\rangle, |5\rangle,...$.

The constant $\delta$ in $\hat{F}(\hat{C}, \hat{N}_0)$ can be fixed by utilizing the commutation relations,
$[\hat{N}_{-}, \hat{\cal N}_{+}]|3\rangle = |3\rangle$ and $[\hat{\cal N}_{-}, \hat{N}_{+}]|3\rangle = |3\rangle$.
From these two relations, we find $\delta = -2$  and fix
${\displaystyle \hat{F}(\hat{C}, \hat{N}_0) = \frac{\hat{N}_0 - 2}{\hat{N}_{-} \hat{N}_{+}}}$.

Finally, one can rescale both the operators $\hat{N}_{+}$ and $\hat{N}_{-}$ simultaneously and evaluate the
commutation relations. For example, let us rescale $\hat{N}_{+}$ and $\hat{N}_{-}$
respectively as $\hat{K}_{+} = \hat{N}_{+} \hat{G}(\hat{C}, \hat{N}_0)$ and
$\hat{K}_{-} = \hat{G}(\hat{C}, \hat{N}_0) \hat{N}_{-}$. The explicit form of $\hat{G}(\hat{C}, \hat{N}_0)$ can be found
by using the commutation relation $[\hat{K}_{-}, \hat{K}_{+}] = \hat{I}$, that is
\begin{eqnarray}
\hat{G}(\hat{C}, \hat{N}_0) \hat{N}_{-} \hat{N}_{+} \hat{G}(\hat{C}, \hat{N}_0) - \hat{N}_{+} \hat{G}^{2}(\hat{C}, \hat{N}_0) \hat{N}_{-} = \hat{I}.
\label{case3}
\end{eqnarray}
Solving Eq. (\ref{case3}) we find $\hat{G}(\hat{C}, \hat{N}_0) = \sqrt{F(\hat{C}, \hat{N}_0)}$.

With this choice of $\hat{G}(\hat{C}, \hat{N}_0)$ we can establish
\begin{eqnarray}
 \mbox{Case: (iii)}\; [\hat{K}_{-}, \hat{K}_{+}]|n\rangle = |n\rangle, \;\; [\hat{K}_{0}, \hat{K}_{-}]|n\rangle = -\hat{K}_{-}|n\rangle,  \;\;\;
[\hat{K}_{0}, \hat{K}_{+}]|n\rangle = \hat{K}_{+}|n\rangle,
\label{lad16}
\end{eqnarray}
where $\hat{K}_{0} = \hat{K}_{+} \hat{K}_{-}$. Here $\hat{K}_0$ serves as a number operator.

We construct coherent and nonlinear coherent states using these three sets of new deformed ladder operators.

\section{Nonlinear coherent states}

\subsection{Non-unitary displacement type operators and nonlinear coherent states}

The transformed operators $\hat{\cal N}_{+}$ and $\hat{\cal N}_{-}$ which satisfy the commutation relations
given in Eqs. (\ref{lad14}) and (\ref{lad15}) help us to define two non-unitary displacement type operators, namely
\begin{eqnarray}
\mbox{Case: (i)}\;\;\;\;\hat{D}(\alpha) &=& e^{\alpha \hat{\cal N}_{+} - \alpha^{*} \hat{N}_{-}},  \\
\label{lad18a}
\mbox{Case: (ii)}\;\;\;\;\hat{\widetilde{D}}(\alpha) &=& e^{\alpha \hat{N}_{+} - \alpha^{*} \hat{\cal N}_{-}}.
\label{lad18}
\end{eqnarray}
By applying these two operators on the lowest energy state $|3\rangle$, given in Eq. (\ref{lad3c}),
after multiplying the displacement operators by $e^{|\alpha|^2 / 2}$, we obtain nonlinear coherent states in the form \cite{Shan, Manko}
\begin{eqnarray}
\mbox{Case: (i)}\;\;\;\;|\alpha, \tilde{f}\rangle &=&  N_{\alpha}\sum^{\infty}_{n=0}\frac{\alpha^{n}}{\sqrt{\widetilde{(n+3)!}}\;\tilde{f}(n+3)!}|n+3\rangle,
\label{lad19b}
\\
\mbox{Case: (ii)}\;\;\;\;\widetilde{|\alpha, \tilde{f}\rangle} &=& \tilde{N}_{\alpha}\sum^{\infty}_{n=0}
                                                         \frac{\alpha^{n}\;\sqrt{\widetilde{(n+3)!}}\;\tilde{f}(n+3)!}{n!} |n+3\rangle,
\label{lad20b}
\end{eqnarray}
respectively. In the above expressions, $\tilde{f}(n + 3)! = f(n+3)f(n+2)f(n+1)...f(4)$,
$\widetilde{(n+3)}! = (n+3)(n+2)(n+1)...4$, and $\tilde{f}(3)! = \tilde{3}! = 1$.
The normalization constant, $N_{\alpha}$,  which can be calculated from the normalization
condition, $\langle \alpha, \tilde{f}|\alpha, \tilde{f}\rangle = 1$, is found to be
\begin{eqnarray}
\mbox{Case: (i)}\;\;\;\;\;N_{\alpha} &=&  \left(\sum^{\infty}_{n=0}\frac{|\alpha|^{2n}}{n!(n+2)!(n+3)!}\right)^{-1/2}.
\label{norm}
\end{eqnarray}
The nonlinear coherent states $|\alpha, \tilde{f}\rangle$ given in Eq. (\ref{lad19b}) coincide with
the one given in our earlier paper \cite{chi2}.

We find that the series given in Eq. (\ref{lad20b}) is a divergent one  and hence
the dual state is not normalizable. From this observation, we conclude that
for the generalized isotonic oscillator one can construct only  nonlinear coherent
states and not their dual counterparts.

\subsection{Unitary operator and coherent states}
In this case the operators, $\hat{K}_{-}, \hat{K}_{+}$ and $\hat{K}_{0}$, act on the states $|n\rangle$ as
\begin{eqnarray}
 \hat{K}_{-}|n\rangle = \sqrt{n-3} |n-1\rangle, \hat{K}_{+}|n\rangle = \sqrt{n-2} |n+1\rangle,
\hat{K}_{0} |n\rangle = (n-3) |n\rangle, \; n = 3, 4, 5, ... . \vspace{0.1cm}
\label{c3op}
\end{eqnarray}

We observe that these operators act, on the states $|3\rangle, |4\rangle, |5\rangle, ...$ of  Eq. (\ref{lad3b}),
in the same way as the annihilation ($\hat{a})$, creation $(\hat{a}^{\dagger})$ and number operators ($\hat{n}$)
act on the photon number states $|0\rangle, |1\rangle, |2\rangle, ...$, of harmonic oscillator.
Using the operators, $\hat{K}_{+}, \hat{K}_{-}$ and $\hat{K}_{0}$, we define an unitary displacement operator, namely
\begin{eqnarray}
\hat{D}(\zeta) = e^{\zeta \hat{K}_{+} - \zeta^{*} \hat{K}_{-}}.
\label{lad22}
\end{eqnarray}
By applying this displacement operator on the lowest energy state, given in Eq. (\ref{lad3c}),
we can obtain the coherent states of the form
\begin{eqnarray}
|\zeta\rangle = \hat{D}(\zeta) |3\rangle = e^{-\frac{|\zeta|^2}{2}}\sum^{\infty}_{n=0}\frac{\zeta^{n}}{\sqrt{n}!}|n+3\rangle.
\label{lad23}
\end{eqnarray}
One may observe that these states have the same form as that of the harmonic oscillator  \cite{How}.
We will discuss the properties exhibited by these coherent states in next section and demonstrate how
these non-Gaussian wave packets minimize the uncertainty relation .


\subsection{Completeness condition}
\label{ssec1}

The utmost requirement of the coherent states is that they should furnish the resolution of unity.
In this sub-section, we investigate whether the nonlinear coherent states
${|\alpha, \tilde{f}} \rangle$ form a complete set of states in the Hilbert space or not.
To establish this we invoke the completeness relation \cite{rii}

\begin{eqnarray}
\frac{1}{\pi} \int \int_\mathbb{C} |\alpha, \tilde{f}\rangle W(|\alpha|^2)\langle \alpha, \tilde{f}| d^2\alpha = \hat{I},
\label{d1}
\end{eqnarray}
where $W(|\alpha|^2)$ is a positive weight function and $\hat{I}$ is an identity operator.
From Eq. (\ref{d1}) we obtain
\begin{eqnarray}
\frac{1}{\pi} \int \int_\mathbb{C} \langle\psi|\alpha, \tilde{f}\rangle W(|\alpha|^2)\langle \alpha, \tilde{f}|\Phi\rangle d^2\alpha = \langle\psi|\hat{I}|\Phi\rangle.
\label{d3}
\end{eqnarray}
Substituting $|\alpha,\tilde{f}\rangle$ and its conjugate (vide Eq. (\ref{lad19b})) in the left hand side of Eq. (\ref{d3}),
we get (which we call $G$)
\begin{eqnarray}
\hspace{-1cm} \qquad G = \frac{1}{\pi} \sum^{\infty}_{m,n = 0}\frac{\langle\psi|n\rangle \langle m|\Phi\rangle}{\sqrt{(n+3)!\;(n+2)!\;n!\;(m+3)!\;(m+2)!\;m!}}\int \int_\mathbb{C} \alpha^{n}{\alpha^{*}}^{m} \tilde{N}^2(|\alpha|^2) W(|\alpha|^2)d^2\alpha.\hspace{0.5cm}
\label{d4}
\end{eqnarray}

Taking $\alpha = re^{i\theta}$, one can separate the real and imaginary parts and obtain
\begin{eqnarray}
\hspace{-1cm} \quad G = \frac{1}{\pi}\sum^{\infty}_{m,n = 0}\frac{\langle\psi|n\rangle\langle m|\Phi\rangle}{\sqrt{(n+3)!\;(n+2)!\;n!\;(m+3)!\;(m+2)!\;m!}} \int^{\infty}_{0} \tilde{N}^{2}(r^2) r^{n+m}\;W(r^2)r dr \nonumber \\ \hspace{2.5cm}\times  \int^{2\pi}_{0} e^{i(n-m)\theta}d\theta.
\label{d5}
\end{eqnarray}
Since the second integral vanishes except $n = m$ we can bring Eq. (\ref{d5}) to the form
\begin{eqnarray}
G &=& \sum^{\infty}_{n=0}\frac{\langle\psi|n\rangle\langle n|\Phi\rangle}{(n+3)!\;(n+2)!\;n!} \int^{\infty}_{0} \tilde{N}^{2}(r^2) r^{2n}W(r^2) 2 r dr.
\label{d6}
\end{eqnarray}
Taking $r^2 = x$, we find
\begin{eqnarray}
G = \sum^{\infty}_{n=0}\frac{\langle \psi|n\rangle\langle n|\Phi\rangle}{(n+3)!\;(n+2)!\;n!} \int^{\infty}_{0} x^{n}\tilde{N}^{2}(x)W(x)dx.
\label{d7}
\end{eqnarray}

Choosing $\tilde{N}^{2}(x) W(x) = G^{4,1}_{2, 5}\left(x \Bigg| \begin{array}{c}
                                                       0, 5\\
                                                       0,2,3,5,0\\
                                                   \end{array}\right) $,
where $G^{m,n}_{p, q}\left(x \Bigg| \begin{array}{c}
                                     a_1, ... a_n, a_{n+1},...a_p\\
                                     b_1, ... b_m, b_{m+1},...b_q\\
                                    \end{array}\right)$, $ m \in N \wedge n \in N \wedge p \in N \wedge q \in N
\wedge m \le q \wedge n \le p$,  is the Meijer $G$-function \cite{book}, the integral on the right hand side in
Eq. (\ref{d7}) can be brought to the form

\begin{equation}
\int^{\infty}_{0} x^{n}\; G^{4,1}_{2, 5}\left(x \Bigg| \begin{array}{c}
                                                       0, 5\\
                                                       0,2,3,5,0\\
                                                   \end{array}\right)  dx
\end{equation}
which can be integrated in terms of gamma functions, namely $\Gamma(n+1)\;\Gamma(n+3)\;\Gamma(n+4)$ \cite{book}.
As a result one gets
\begin{eqnarray}
G &=& \sum^{\infty}_{n=0}\langle\psi|n + 3\rangle\langle n + 3|\Phi \rangle.
\label{d9}
\end{eqnarray}
We mention here that the nonlinear coherent states contain the states with photon number greater than or equal to three.
As a consequence Eq. (\ref{d9}) turns out to be a projector in  which the state $n = 0$ is excluded.

As we mentioned earlier, the coherent states of Case (iii), given in Eq. (\ref{lad23}), are in the same form as
that of harmonic oscillator coherent states.  This type of coherent states satisfy
the following two Hilbert space properties as that of harmonic oscillator coherent states  \cite{How}, namely 
(i) overlap between two different coherent states and  (ii) linear independency among the
finite number of distinct coherent states. 
As far as the completeness condition is concerned
these states form an orthogonal projector in the subspace ${\cal H'}$ only (vide Eq. (\ref{d9})).

\section{Non-classical properties}
\label{prop}
One can analyze the classical and non-classical nature of the
states by investigating photon statistical properties of the states. The coherent states
exhibit Poissonian probability distribution and are said to be closest to
the classical description of a radiation field. In contrast to coherent states, the non-classical states exhibit
non-classical properties such as sub-Poissonian and super-Poissonian statistics, quadrature squeezing and higher order
squeezing.

\subsection{$A_3$-parameter}

In this sub-section, we investigate the parameter $A_3$
associated with the nonlinear coherent states (Eq. (\ref{lad19b})). This parameter
was introduced as a counterpart to the Mandel's parameter $Q$ by Agarwal and Tara in Ref. \onlinecite{agarwal1}.
It was also recently studied for the newly introduced $\beta$-nonlinear coherent states \cite{tavassoly_beta}.
To test the non-classical character of the field even it does not exhibit the squeezing and
sub-Poissonian statistics, the Mandel's parameter $Q$ is generalized to a quantity $m^{(n)}$ formed
from the moments of Glauber-Sudarshan  function ($P$), $\hat{m}_n = \hat{b}^{\dagger^n} {\hat{b}}^n$,
where $\hat{b}$ and $\hat{b}^{\dagger}$ are annihilation and creation operators of the harmonic oscillator.
To normalize $m^{(n)}$ another quantity $\mu^{(n)}$, where $\hat{\mu}^{(n)} = (\hat{b}^{\dagger} \hat{b})^n$,
which is formed from the moments of number distribution has also been introduced.
The normalized quantity with $n = 3$ obtained from $m^{(n)}$ and $\mu^{(n)}$ is termed as parameter $A_3$
(see below Eq. (\ref{A3})).  The parameter $A_3$ can be calculated from the
expression \cite{agarwal1},

\begin{eqnarray}
A_{3} = \frac{\det m^{(3)}}{\det\mu^{(3)} - \det m^{(3)}},
\label{A3}
\end{eqnarray}
where,
\begin{eqnarray}
     m^{(3)}= \left(\begin{array}{ccc}
    1 & m_1 & m_2 \\
    m_1 & m_2 & m_3 \\
    m_2& m_3 & m_4\\
    \end{array} \right) \;\;\;\mbox{and}
\;\;\;\;\;&
  \mu^{(3)}= \left(
  \begin{array}{ccc}
   1 & \mu_1 & \mu_2 \\
    \mu_1 & \mu_2 & \mu_3 \\
    \mu_2& \mu_3 & \mu_4\\
    \end{array}
   \right).
\end{eqnarray}

\begin{figure}[t]
\begin{center}
\hspace{-7.3cm}\includegraphics[width=0.9\linewidth]{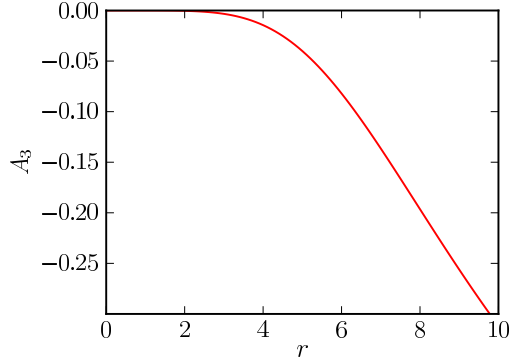}
\end{center}
\vspace{-6.5cm}
\caption{The plots of the parameter $A_3$ which is calculated with respect to
nonlinear coherent states Eq. (\ref{lad19b}).}
\label{a3_nlcs}
\end{figure}
In the above, $m_j = \langle\hat{m}_j\rangle= \langle \hat{K}^{j}_{+} \hat{K}^j_{-} \rangle $ and 
$\mu_j = \langle \hat{\mu}_j \rangle= \langle(\hat{K}_{+} \hat{K}_{-})^j \rangle$,\;$j = 1, 2, 3, 4 $.
For the coherent and vacuum states $\det m^{(3)}=0$ and for a Fock state $\det m^{(3)}= -1$ and $\det\mu^{(3)} = 0$.
For the non-classical states $\det m^{(3)} < 0$ and since $\det\mu^{(3)} > 0$, it follows that parameter $A_3$
lies between 0 and -1.

To calculate the parameter $A_3$, we  evaluate ${m}_j$'s and
${\mu}_j$'s, $j=1,2,3,4$, for the nonlinear coherent states $|\alpha, \tilde{f}\rangle$ as
\begin{eqnarray}
 m_j  =  N^2_{\alpha} \sum^{\infty}_{n=j}\frac{|\alpha|^{2n}}{(n-j)!(n+2)!(n+3)!}.
\label{m3}
\end{eqnarray}

In a similar way one can obtain the expression for  ${\mu}_j$ in the form
\begin{eqnarray}
 \mu_j  = N^2_{\alpha} \sum^{\infty}_{n=1}\frac{|\alpha|^{2n}\; n^{j-1}}{(n-j)!(n+2)!(n+3)!}.
\label{mu3}
\end{eqnarray}

From the expectation values we work out the parameter $A_3$ for the nonlinear
coherent states $|\alpha, \tilde{f}\rangle$ numerically.  We then plot the results in Fig. \ref{a3_nlcs}
where we draw the parameter $A_3$ against $r\;(= |\alpha|)$. The parameter $A_3$
lies between 0 to -1 for the nonlinear coherent states given in Eq. (\ref{lad19b}).

\subsection{Quadrature squeezing}
\label{sq}
The non-classical nature of a quantum state can also be characterized by
examining the degree of squeezing \cite{walls} it possesses.
Since $\hat{K}_{-}$ and $\hat{K}_{+}$ act as annihilation $(\hat{a})$ and
creation $(\hat{a}^{\dagger})$ operators for the system (\ref{lad3b}), we define two
Hermitian operators, $\hat{x}$ and $\hat{p}$, namely \cite{walls}.
\begin{eqnarray}
\hat{x} = \frac{1}{\sqrt{2}}(\hat{K}_{+} + \hat{K}_{-}), \qquad
\hat{p} = \frac{i}{\sqrt{2}}(\hat{K}_{+}- \hat{K}_{-}).
\label{ass3}
\end{eqnarray}

Then the Heisenberg uncertainty relation holds,
$\left(\Delta \hat{x} \right)^2\left(\Delta \hat{p} \right)^2 \ge \frac{1}{4}$,
where $\Delta \hat{x}$ and $\Delta \hat{p}$ denote uncertainties in $\hat{x}$ and $\hat{p}$
respectively. In general, a state is squeezed, if any of the following conditions holds:
$(\Delta \hat{x})^2 < \frac{1}{2}$ or $(\Delta \hat{p})^2 < \frac{1}{2}$.  Using
the expressions given in (\ref{ass3}), the squeezing conditions can be transformed to the following inequalities,
that is
\begin{eqnarray}
\quad I_{1} &=& \langle {\hat{K}_{-}}^2 \rangle + \langle {\hat{K}_{+}}^2 \rangle - \langle {\hat{K}_{-}} \rangle^{2} - \langle {\hat{K}_{+}} \rangle^{2} - 2 \langle \hat{K}_{-} \rangle \langle \hat{K}_{+} \rangle + 2\langle \hat{K}_{+} \hat{K}_{-} \rangle < 0,
\label{id1}\\
\quad I_{2} &=& -\langle {\hat{K}_{-}}^2 \rangle - \langle {\hat{K}_{+}}^2 \rangle + \langle {\hat{K}_{-}} \rangle^{2} + \langle {\hat{K}_{+}} \rangle^{2} - 2 \langle \hat{K}_{-} \rangle \langle \hat{K}_{+} \rangle + 2\langle \hat{K}_{+} \hat{K}_{-} \rangle < 0,
\label{id2}
\end{eqnarray}
where the expectation values are calculated with respect to the nonlinear coherent states
$|\alpha, \tilde{f}\rangle$ for which the squeezing property has to be examined.

\begin{figure}[ht]
\centering
\includegraphics[width=0.9\linewidth]{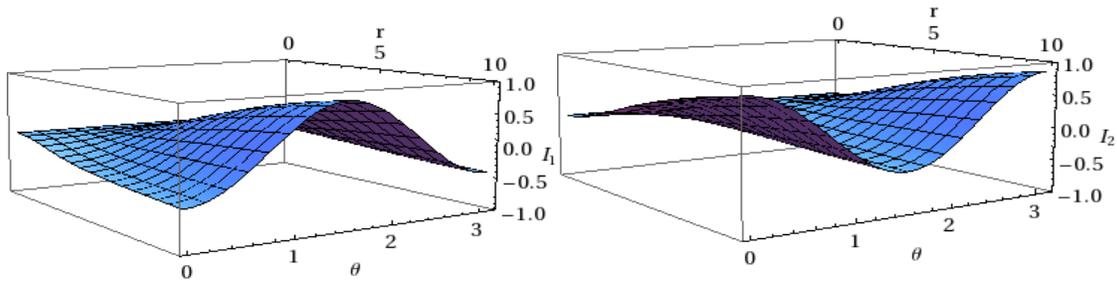}
\vspace{-0.7cm}
\caption{The plots of $I_1$ and $I_2$ which are calculated with respect to
nonlinear coherent states (\ref{lad19b}) for $n_{max} = 200$.}
\label{c1_cs_id_12}
\end{figure}
The identities given in Eqs. (\ref{id1}) and (\ref{id2}) are evaluated numerically and
plotted in Fig. \ref{c1_cs_id_12} with $\alpha = r e^{i\;\theta}$. From the figure \ref{c1_cs_id_12} we
observe that the identities  given in Eqs. (\ref{id1}) and (\ref{id2}) for the nonlinear coherent states
$|\alpha, \tilde{f}\rangle$, satisfying the uncertainty relation
show small oscillations in $I_1$ and $I_2$. These two quantities, $I_1$ and $I_2$, oscillate out of phase
$\pi$ with each other. This in turn reveals that a small degree of squeezing can be observed in both
the quadratures, $\hat{x}$ and $\hat{p}$, at different values of $\theta$.

As far as Case (iii) is concerned, since the photon statistical properties have the classical nature
they do not admit any squeezing property. However, in the following,
we demonstrate that these coherent states are minimum uncertainty states.
Recalling the definition and using Eq. (\ref{ass3}) we find
\begin{eqnarray}
\qquad \left(\Delta \hat{x} \right)^2 \left(\Delta \hat{p} \right)^2 = \frac{1}{4}
\label{j38}
\end{eqnarray}
from which we deduce
\begin{eqnarray}
\qquad \left(\Delta \hat{x} \right)^2 = \left(\Delta \hat{p} \right)^2 = \frac{1}{2}.
\label{j39}
\end{eqnarray}

Thus, the coherent states $|\zeta\rangle$ minimize the uncertainty relation but do not exhibit the squeezing property.

\subsection{Amplitude-squared squeezing}
The amplitude-squared squeezing, which was introduced by Hillery \cite{hillery}, involves two operators
which represent the real and imaginary parts of the square of the amplitude of a radiation field.

To investigate the amplitude-squared squeezing effect
for the Case (i), we introduce again two Hermitian operators $\hat{X}$ and $\hat{P}$
from $\hat{K}_{+}$ and $\hat{K}_{-}$ respectively of the form
\begin{eqnarray}
\hat{X} = \frac{1}{\sqrt{2}}(\hat{K}^2_{+} + \hat{K}^2_{-}), \qquad
\hat{P} = \frac{i}{\sqrt{2}}(\hat{K}^2_{+}- \hat{K}^2_{-}).
\label{as3}
\end{eqnarray}
Here $\hat{X}$ and $\hat{P}$ are the operators corresponding to the real and imaginary parts
of the square of the complex amplitude of a radiation field. Heisenberg uncertainty relation of these two pairs of
conjugate operators is then given by $(\Delta \hat{X})^2 (\Delta \hat{P})^2 \ge -\frac{1}{4}\langle[\hat{X}, \hat{P}]\rangle^2$.
For the nonlinear coherent states (\ref{lad19b}), we find
that the states satisfy  the uncertainty relation.
By following the steps given in quadrature squeezing,
one can straightforwardly derive conditions for the amplitude-squared squeezing. The conditions read
\begin{eqnarray}
 \qquad I_{3}  = \frac{1}{4} \left(\langle {\hat{K}_{-}}^4 \rangle + \langle {\hat{K}_{+}}^4\rangle
                     - \langle {\hat{K}_{-}}^2 \rangle^{2} - \langle {\hat{K}_{+}}^2 \rangle^{2}
                     - 2 \langle {\hat{K}_{-}}^2 \rangle \langle {\hat{K}_{+}}^2 \rangle
                      +\langle {\hat{K}_{+}}^2 {\hat{K}_{-}}^2 \rangle + \langle {\hat{K}_{-}}^2 {\hat{K}_{+}}^2 \rangle \right) \nonumber
         \\ - \langle {\hat{K}_{+}} {\hat{K}_{-}} \rangle - \frac{1}{2} < 0,
\label{id3}\\
 \qquad I_{4} = \frac{1}{4} \left(-\langle {\hat{K}_{-}}^4 \rangle - \langle {\hat{K}_{+}}^4\rangle
                   +\langle {\hat{K}_{-}}^2 \rangle^{2}+ \langle {\hat{K}_{+}}^2 \rangle^{2}
                   - 2 \langle {\hat{K}_{-}}^2 \rangle \langle {\hat{K}_{+}}^2 \rangle
                      +\langle {\hat{K}_{+}}^2 {\hat{K}_{-}}^2 \rangle + \langle {\hat{K}_{-}}^2 {\hat{K}_{+}}^2 \rangle \right) \nonumber
         \\ - \langle {\hat{K}_{+}} {\hat{K}_{-}} \rangle - \frac{1}{2} < 0,
\label{id4}
\end{eqnarray}
where the expectation values are calculated with respect to the nonlinear coherent states
$|\alpha, \tilde{f}\rangle$ for which the squeezing property has to be examined.
\begin{figure}[ht]
\centering
\includegraphics[width=0.9\linewidth]{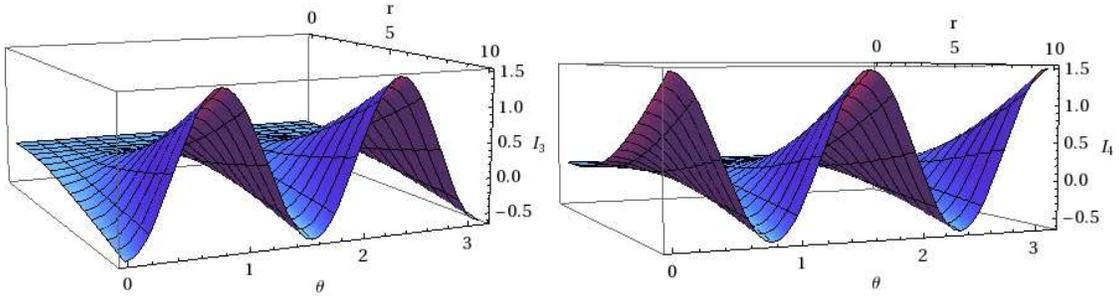}
\vspace{-0.7cm}
\caption{The plots of $I_3$ and $I_4$ which are calculated with respect to
nonlinear coherent states (\ref{lad19b}) for $n_{max} = 200$.}
\label{c1_cs_id_34}
\end{figure}

We evaluate these two identities, in Eqs. (\ref{id3}) and (\ref{id4}), numerically and
plot the results in Fig.  \ref{c1_cs_id_34}.
The identities $I_3$ and $I_4$ also vary in an oscillatory manner.
For certain values of $r$ and $\theta$ when one of the identities $(I_3)$ is positive the
other identity $(I_4)$ becomes negative, which shows the squeezing in one of the operators
$\hat{X}$ and $\hat{P}$. The results confirm the non-classical behaviour
exhibited by nonlinear coherent states (\ref{lad19b}).

In  Case (iii), from the definition of coherent states $\hat{K}_{-}|\zeta\rangle = \zeta |\zeta\rangle$, we find
\begin{eqnarray}
\quad \langle\zeta|\hat{X}|\zeta\rangle &=& \frac{1}{2}(\zeta^2+{\zeta^*}^{2}),\qquad \;\;
\langle\zeta|\hat{X}^2|\zeta\rangle =  \frac{1}{2}(\zeta^2+{\zeta^*}^{2})^2+\left(2 |\zeta|^2 + 1\right),
\label{rel1}\\
\quad \langle\zeta|\hat{P}|\zeta\rangle &=& \frac{-i}{\sqrt{2}}({\zeta^*}^2-\zeta^2),\qquad
\langle\zeta|\hat{P}^2|\zeta\rangle = -\frac{1}{2}({\zeta^*}^{2}- \zeta^2)^2+\left(2|\zeta|^2 + 1\right).
\label{rel2}
\end{eqnarray}

Using the expressions (\ref{rel1}) and (\ref{rel2}) we can show the
coherent states $|\zeta\rangle$ minimize the uncertainty relation as
\begin{eqnarray}
\left(\Delta \hat{X} \right)^2\left(\Delta \hat{P} \right)^2 = \left(2|\zeta|^2 + 1\right)^2
\label{amp1}
\end{eqnarray}
with
\begin{eqnarray}
\left(\Delta \hat{X} \right)^2 = \left(\Delta \hat{P} \right)^2 = \left(2|\zeta|^2 + 1\right).
\label{amp2}
\end{eqnarray}
Hence, the coherent states minimize the uncertainty relation (vide Eq. (\ref{amp1}))
and does not exhibit amplitude-squared squeezing property.

\subsection{Photon statistical properties}
For the nonlinear coherent states (\ref{lad19b}), Case (i), the photon statistical
properties, such as photon number probability distribution,
Mandel's parameter and the second order correlation function, were  studied in Ref. \onlinecite{chi2}.
So we proceed to the Case (iii) straightaway.

As far as Case (iii) is concerned, we find photon number distribution is of the form
\begin{eqnarray}
P(n) = |\langle n+ 3|\zeta\rangle|^2 = \frac{|\zeta|^{2 n}}{n!}e^{-|\zeta|^2}\label{ps3}
\label{c3_p3}
\end{eqnarray}
which is nothing but a Poisson distribution. In the above, we have defined $P(n)$ in such a
way that it calculates the probability distribution of the photon number which starts from the
state $|3\rangle$. This is due to the fact that in the present problem the Hilbert space
consists of states $|0\rangle, |3\rangle, |4\rangle, ...$ has been split up into two invariant
subspaces and we consider only the sub-Hilbert space spanned by the eigenstates $|3\rangle, |4\rangle, |5\rangle, ...$.
Because of this, the sub-Hilbert space contains the states
$|3\rangle, |4\rangle, |5\rangle,...$ should be considered as $|\tilde{0}\rangle,
|\tilde{1}\rangle, |\tilde{2}\rangle,...$ with eigenvalues $0, 1, 2$,...
respectively. We have defined the operators $\hat{K}_{-}$ and $\hat{K}_{+}$, where $\hat{K}_{+}$ is
the adjoint of $\hat{K}_{-}$, in  such a way that they satisfy the commutation relation
$[\hat{K}_{-}, \hat{K}_{+}]|n\rangle = (n-3)|n\rangle$, $n = 3, 4, 5, ...$ which can be 
rewritten as
\begin{equation}
[\hat{K}_{-}, \hat{K}_{+}]|\tilde{n}\rangle = n |\tilde{n}\rangle, \quad n = 0, 1, 2, ....\quad
\mbox{where}\; |\tilde{n}\rangle := |n+3\rangle.
\end{equation}
The operators $\hat{K}_{-}$ and $\hat{K}_{+}$ factorize the operator $\hat{K}_0$ as $\hat{K}_{0} =
\hat{K}_{+}\hat{K}_{-}$. The operator $\hat{K}_0$ can be interpreted as the number operator 
since it satisfies the relation $\hat{K}_{0}|\tilde{n}\rangle = n |\tilde{n}\rangle $ and
possesses positive and zero eigenvalues \cite{schiff}. By 
excluding the ground state $|0\rangle$, the Hamiltonian $\tilde{H}$ can now be defined in the sub-Hilbert space
consists of the states $|3\rangle, |4\rangle, |5\rangle, ...$
that is $|\tilde{0}\rangle, |\tilde{1}\rangle, |\tilde{2}\rangle,...$ as
\begin{equation}
\tilde{H} |\tilde{n}\rangle = \tilde{e}_{n} |\tilde{n}\rangle , \qquad n = 0, 1, 2, ...
\end{equation}
with energy eigenvalues $\tilde{e}_{n} = n \hbar \omega$. Hence the values $n$ represent the number of photons. 
In this respect the number operator $\hat{K}_{0}$ determines the number
of photons in each mode if each mode of the electomagnetic field is considered to be an oscillator  
described by the Hamiltonian $\tilde{H}$.

\vspace{0.5cm}
\begin{figure}[ht]
\centering
\includegraphics[width=0.9\linewidth]{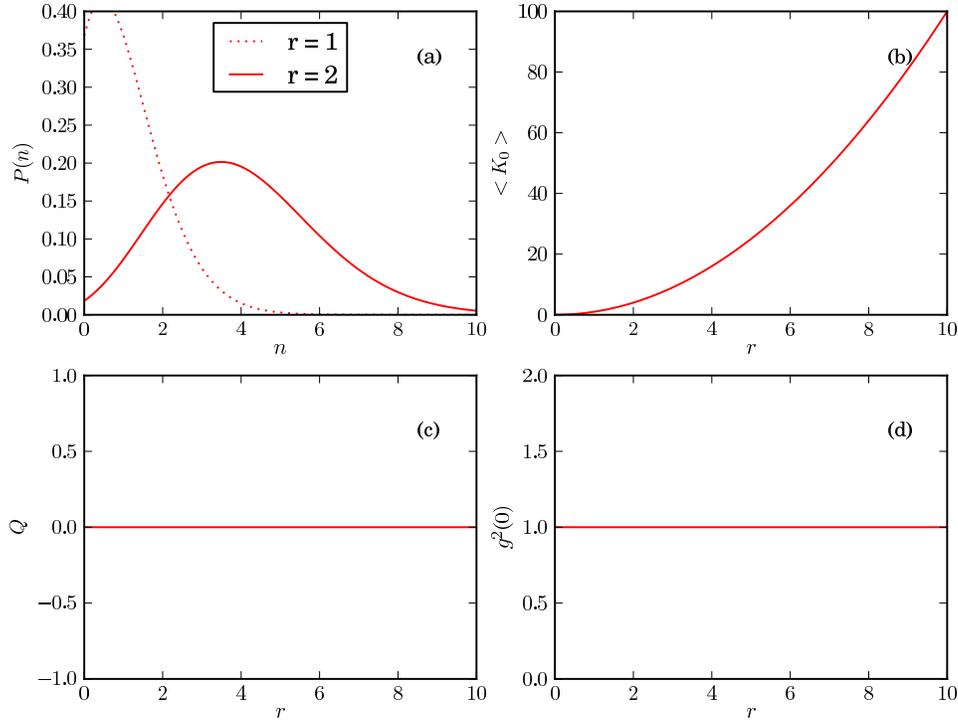}
\vspace{-0.5cm}
\caption{The plots of (a) photon number distribution $P(n)$, (b) mean photon number $\langle \hat{K_0} \rangle$,
(c) Mandel's parameter $Q$  and (d) the second order correlation function $g^{2}(0)$ of coherent states
(\ref{lad23}).}
\label{cs_c3_prop}
\end{figure}
We calculate Mandel's parameter and the second order correlation function for
this case as well, with the help of the operator $\hat{K}_0$, by using the expressions \cite{Mandel, Mah, Ant, Paul}
\begin{eqnarray}
Q &=& \frac{\langle \hat{K}^{2}_{0} \rangle}{\langle \hat{K}_{0} \rangle} - \langle \hat{K}_{0} \rangle -1,
\label{q}\\
g^{2}(0) &=& \frac{\langle \hat{K}^{2}_{0} \rangle - \langle \hat{K}_{0} \rangle}{\langle \hat{K}_{0} \rangle^2}.
\label{g}
\end{eqnarray}

The expectation values of $\langle \hat{K}_0 \rangle $ and $\langle \hat{K}^2_0 \rangle $ are turned out to be
\begin{eqnarray}
\langle \hat{K}_0 \rangle = \sum^{\infty}_{n=1} \frac{|\zeta|^{2 n}}{(n-1)!},  \;\;\;
\langle \hat{K}^2_0 \rangle = \sum^{\infty}_{n=1} \frac{|\zeta|^{2 n}n}{(n-1)!},
\label{expc3}
\end{eqnarray}
where $\langle \hat{K}_0 \rangle$ is the average number of photons. We draw the average number of photons,
$\langle \hat{K}_0 \rangle$, for different values of $r$ in Fig.  \ref{cs_c3_prop}(b)
which shows the linear dependence of $\langle \hat{K}_0 \rangle$  on $r$.
Substituting Eq. (\ref{expc3}) in Eqs. (\ref{q}) and (\ref{g}), we find $Q = 0$
and the second order correlation function $g^{2}(0) = 1$.
The results are given in Figs.  \ref{cs_c3_prop}(c)
and \ref{cs_c3_prop}(d) respectively which once again confirm the Poissonian nature of the coherent states (Eq. (\ref{lad23}))
for all values of $r$.

\section{Quadrature distribution  and quasi-probability functions}
\label{quasi}

\subsection{Phase-parameterized field strength distribution}

In this sub-section, we calculate phase-parameterized field strength
distribution of the nonlinear coherent states $|\alpha, \tilde{f}\rangle$,
which is defined by
\begin{eqnarray}
P(x, \phi) = |\langle x, \phi|\alpha, \tilde{f}\rangle|^{2},
\label{para1}
\end{eqnarray}
where $|x,\phi\rangle$ is the eigenstate of the quadrature component
$\hat{x}(\phi) = \frac{1}{\sqrt{2}}\left(e^{- i \phi} \hat{K}_{-} + e^{i \phi} \hat{K}_{+} \right)$, that is
\begin{equation}
\hat{x}(\phi)|x,\phi\rangle = x|x,\phi\rangle,
\label{xeq}
\end{equation}
which can be expressed in photon number basis as \cite{obada}
\begin{eqnarray}
|x,\phi\rangle = \frac{e^{\frac{-x^2}{2}}}{\pi^{\frac{1}{4}}} \sum^{\infty}_{n=0} \frac{H_n(x) e^{i n \phi }}{\sqrt{2^n n!}}|n+3\rangle,
\label{xphi}
\end{eqnarray}
where $H_{n}(x)$ is the Hermite polynomial. Considering $\alpha = r e^{i \theta}$ and
using Eqs. (\ref{lad19b}) and (\ref{xphi})   in Eq. (\ref{para1}),
we arrive at
\begin{eqnarray}
\quad P(x, \phi) =\frac{N^{2}_{\alpha} e^{-x^2}}{\sqrt{\pi}} \sum^{\infty}_{n,m = 0} \frac{r^{n+m}\cos[(n-m)(\theta - \phi)]
H_{n}(x)H_{m}(x)}{n!\;m!\sqrt{2^{n+m} (n+2)! (m+2)! (n+3)!  (m+3)!}}.
\label{pa_nlcs}
\end{eqnarray}

\begin{figure}[!ht]
\vspace{-0.5cm}
\begin{center}
\includegraphics[width=0.6\linewidth]{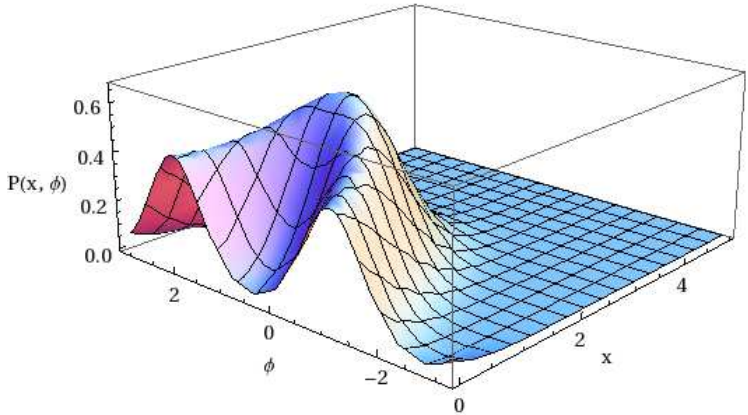}
\end{center}
\vspace{-0.9cm}
\caption{The plot of $P(x,\phi)$ for nonlinear coherent states  (\ref{lad19b}) for $ r = 5$ and $\theta = 0.5$.
}\label{pa_nlcs_fig}
\end{figure}

From the expression (\ref{pa_nlcs}), we calculate the quadrature distribution numerically.
We plot the function $P(x, \phi)$ in Fig. \ref{pa_nlcs_fig} for $r = 5$ and $\theta =  0.5$.
The results match with the one discussed in Ref. \onlinecite{obada}. From
Fig. \ref{pa_nlcs_fig}, we observe that $P(x,\phi)$ is symmetric around $\phi = 0$. Further
we observe that at $x = 0$  a two-peak shape appears around $\phi = \pm \frac{\pi}{2}$
in the distribution function and it converges as $x$ increases. When $x > 2$, this two-peak shape
submerges and a single broad peak appears as $\phi$ gets closer to $0$.
Finally, all phase information disappears when $x$ exceeds the value $2.8$. We observe that any change in $\alpha$ makes
a difference in the two-peak shape.

\subsection{$s$-parameterized quasi-probability function}
In this sub-section, we study three different quasi-probability distribution functions,
namely (i) Wigner function $(W(x, p))$, (ii) Glauber - Sudarshan function $(P)$ and
(iii) Husimi function $(Q(x, p))$  of the nonlinear coherent states (\ref{lad19b})
constructed for the system given in Eq. (\ref{lad3b}). The Wigner function describes the state of a quantum system in
phase space in the same fashion as the probability distribution function (non-negative by definition),
which characterizes a classical system.
The Wigner function, which was introduced as quantum corrections in classical statistical
mechanics, normally takes negative values in certain domains of phase space so
that it cannot be interpreted as a classical distribution function which is non-negative
by necessity \cite{wigner, vbook}. Hence, the negativity of the Wigner function is indeed
a good indication of the highly non-classical character exhibited by  the state.
The Glauber-Sudarshan function $(P)$  and the Husimi function ($Q$) are used
to express quantum mechanical expectation values as classical averages, with the respective
distribution function as the weight function. For some states, Glauber-Sudarshan $P$-function is in fact
so strongly singular that it is not even a tempered distribution. Hence the singularity of the Glauber-Sudarshan function
represents the non-classical character of the quantum sates \cite{vbook, gs}.
In contradiction to other two distribution functions, the Husimi Q-function is well behaved and
non-negative. Moreover it is simply expressed by the coherent expectation
value of the field density matrix ${\displaystyle Q(x,p) = \frac{1}{\pi}\langle\zeta|\hat{\rho}|\zeta\rangle}$
and has been widely adopted to describe field dynamics in situations where the density matrix $\hat{\rho}$
can be easily computed \cite{agarwal}. The quasi-probability distribution function is, in fact,
more general than that of the above said three distribution functions. Cahill and Glauber have
generalized the concept of quasi-probability distribution function by introducing
the $s$-parameterized function with $s$ being a continuous variable. This generalized
function interpolates the Glauber-Sudarshan $P$-function for $s$ = 1, Wigner function $W$ for $s = 0$,
and Husimi $Q$-function for $s = -1$ \cite{s_para}.

The $s$-parameterized quasi-probability distribution function, $F(z, s)$, is the
Fourier transformation of the $s$-parameterized characteristic function
\begin{eqnarray}
F(z, s) = \frac{1}{\pi^2}\int C(\lambda, s)\; e^{(\lambda^* z - \lambda z^*)} d^{2} \lambda,
\label{spara}
\end{eqnarray}
where
\begin{eqnarray}
C(\lambda, s) = \mbox{Tr}[\hat{\rho} \hat{D}(\lambda)] \exp{\left[\frac{s}{2} |\lambda|^2\right]}
\label{char}
\end{eqnarray}
is the $s$-parameterized characteristic function \cite{obada}, $\hat{D}(\lambda)$ is
the displacement operator and $s$ is the ordering parameter.
To study quasi-probability distribution of the nonlinear coherent states
of Eq. (\ref{lad3b}), we only consider the unitary displacement operator
$\hat{D}(\lambda)$  with $\hat{K}_{-}$ and $\hat{K}_{+}$, that is
$\hat{D}(\lambda) = \exp{(\lambda \hat{K}_{+} - \lambda^{*} \hat{K}_{-})}$,
since $\hat{K}_{-}$ and $\hat{K}_{+}$  act as annihilation $(\hat{a})$ and
creation ($\hat{a}^{\dagger}$) operators for the system (\ref{lad3b}).
\vspace{0.1cm}
\begin{figure}[ht]
\centering
\includegraphics[width=0.95\linewidth]{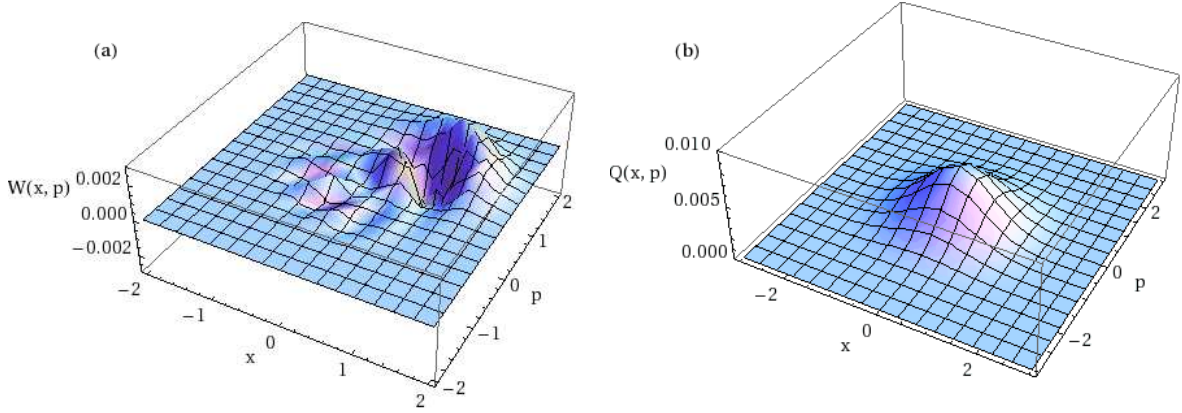}
\vspace{-0.4cm}
\caption{The plots of Wigner function $W(x,p)$ and Husimi $Q(x,p)$ function corresponding to
nonlinear coherent states (\ref{lad19b}) for $\alpha = 10 + i10$.} \label{cs_c1_wig}
\end{figure}

For the nonlinear coherent states $|\alpha, \tilde{f}\rangle$, we find
\begin{eqnarray}
\mbox{Tr}[\hat{\rho} \hat{D}(\lambda)] = \sum^{\infty}_{n,m=0} B_{n,m} \langle m+3|D(\lambda)|n+3 \rangle,
\label{tr}
\end{eqnarray}
where
\begin{eqnarray}
B_{n,m} =   N^2_{\alpha} \sum^{\infty}_{n,m=0} \frac{{\alpha^{*}}^{m} \alpha^{n}}{\sqrt{n!\;m!(n+2)!(m+2)!(n+3)!(m+3)!}}.
\label{Bnm1}
\end{eqnarray}
We find $\langle m+3|\hat{D}(\lambda)|n+3 \rangle$ is to be \cite{s_para, barn}
\begin{eqnarray}
\langle m+3|\hat{D}(\lambda)|n+3 \rangle = \exp{\left[-\frac{|\lambda|^2}{2}\right]}\sqrt{\frac{n!}{m!}} \lambda^{m-n}L^{m-n}_{n}(|\lambda|^2),
\label{disp}
\end{eqnarray}
where $L^{m-n}_n$ is an associated Laguerre polynomial.
Using Eqs. (\ref{disp}) and  (\ref{char}) in  Eq. (\ref{spara}), one can evaluate the function \cite{obada, s_para}
\begin{eqnarray}
 \;\;F(z, s) = \frac{2\exp{\left[\frac{2}{(s-1)}|z|^2\right]}}{\pi(1-s)}\sum^{\infty}_{n,m=0} B_{n,m} (-2 z)^{m-n} \sqrt{\frac{n!}{m!}} \frac{(s+1)^n}{(s-1)^m}L^{m-n}_{n}\left(\frac{4\;|z|^2}{1-s^2} \right). \qquad
\label{spara10}
\end{eqnarray}

Formula (\ref{spara10}) yields the Glauber-Sudarshan $P$-function for $s = 1$, the Wigner $W$-function
for $s = 0$ and the Husimi function for $s = -1$.

Using Eq. (\ref{Bnm1}) in Eq. (\ref{spara10}), we can calculate Wigner function ($s = 0$)
for the nonlinear coherent states $|\alpha, \tilde{f}\rangle$ numerically.
The results are plotted in Figs. \ref{cs_c1_wig}(a).
In Fig. \ref{cs_c1_wig}(a), we consider $\alpha = 10 + i 10$ with $z = x + i p$.
The Wigner function $W(x,p)$ possesses negative values for the nonlinear coherent states $|\alpha, \tilde{f} \rangle$.
This in turn confirms the non-classical nature of the nonlinear coherent
states given in Eq. (\ref{lad19b}). We observe that any slight change in $\alpha$ also causes a
change in the negativity of Wigner distribution function.

Proceeding further, substituting the expressions (\ref{Bnm1}) in Eq. (\ref{spara10}) and considering $z = x + ip$,
we obtain the following formula for the Husimi function ($s = -1$) for $|\alpha, \tilde{f}\rangle$,
that is
\begin{eqnarray}
Q(x, p) = \frac{N^2_{\alpha} }{\pi } e^{-(x^2 + p^2)}.
\label{hus1}
\end{eqnarray}
The function $Q(x, p)$ given in Eq. (\ref{hus1}) is evaluated and plotted in Fig. \ref{cs_c1_wig}(b) for
$\alpha = 10 + i10$. The Fig. \ref{cs_c1_wig}(b) shows that
the distribution is Gaussian for both the nonlinear coherent states, which in turn confirm
that the Husimi function is a smoothed Wigner function. Here also we observe
that Husimi function is sensitive to any change in  $\alpha$.

As mentioned earlier, the Glauber-Sudarshan $P$-function can be deduced from
$s$-parameterized function by fixing $s = 1$ in Eq. (\ref{spara10}), that is
\begin{eqnarray}
P(z) = F(z,1).
\label{def1}
\end{eqnarray}
Since the term $(1-s)$ appears in the denominator the $P$-function becomes infinite for all quantum states.
Hence, the $P$-function for a given density operator $\hat{\rho}$ is alternatively defined to be
\begin{eqnarray}
P(z) = \lim_{s\to1-} F(z, s),
\end{eqnarray}
which can make the function $P(z)$  either as a well behaved function or a tempered distribution \cite{s_para}.
By considering $n = m$ in Eq. (\ref{spara10}), we find
\begin{eqnarray}
P(z) =  \frac{1}{\pi}\sum^{\infty}_{n=0} B_{n,n} (-1)^n e^{|z|^2} \frac{d^n}{d(|z|^2)^n}\delta(|z|^2),
\label{gsre}
\end{eqnarray}
where $\delta(|z|^2)$ is the Dirac-delta function. The expression (\ref{gsre}) shows
that $P(z)$ is highly singular and confirms the non-classical nature of
the nonlinear coherent states $|\alpha, \tilde{f}\rangle$.

\section{Conclusion}
In this paper, we have pointed out the non-existence of dual states for the
generalized isotonic oscillator (\ref{lad1}) with ${ a^2 =  \frac{\hbar}{2 m_0 \omega}}$
and ${ g_a = \frac{2 \hbar^2}{m_0}}$. To show this result, we have obtained the Heisenberg algebra by
transforming the deformed ladder operators suitably. We have shown that these transformations can be
chosen in three different ways. Once the Heisenberg algebra has been identified, we have unambiguously defined
the displacement type  operators in all three cases.
In the first two cases, the displacement operators are of non-unitary type whereas in the
third case it is an unitary one. The non-unitary displacement type operator $\hat{D}(\alpha)$
acts on the lowest energy state $|3\rangle$ of system (\ref{lad3b})
yields the nonlinear coherent states whereas $\hat{\widetilde{D}}(\alpha)$ fails to  produce their dual pair
and the unitary displacement operator provides only the harmonic oscillator coherent states.
The non-classical nature of the nonlinear coherent states has been confirmed through the evaluation of
the parameter $A_3$. Further, we have demonstrated that the nonlinear coherent states
possess certain other non-classical properties as well, namely quadrature and amplitude-squared squeezing. In addition to the
above, we have analyzed the quadrature distribution  and $s$-parameterized quasi-probability function for the
nonlinear coherent states  and confirmed the non-classical nature, exhibited by these states. Finally,
we have derived analytical expressions for the $P$-function, $Q$-function and the Wigner function for the
nonlinear coherent states. As far as the harmonic oscillator coherent states are concerned
our investigations confirmed that they possess
Poissonian distribution only. We have also demonstrated that these coherent states minimize the uncertainty relation.
In addition to the above, we have also discussed some applications and possible extensions of the present work.

\end{document}